\documentclass[12pt,onecolumn,draftcls]{IEEEtran} 

\usepackage{graphicx,subfigure,slashbox}
\usepackage{amsmath,amsfonts}
\usepackage{color}
\usepackage{url}
\usepackage{cite}

\newtheorem{theorem}{Theorem}[section]

\newtheorem{corollary}{Corollary}[section]
\newtheorem{proposition}[theorem]{Proposition}

\newcommand{\qstar}{{q^*}}

\newcommand{\etc}{{\em etc.}~}
\newcommand{\ie}{{\em i.e.},~}
\newcommand{\eg}{{\em e.g.},~}
\newcommand{\cf}{{\em cf.}~}

\newcommand{\uq}{\underline{q}}

\newcommand{\uone}{\underline{1}}
\newcommand{\e}{\mbox{e}}

\newcommand{\beqa}{\begin{eqnarray*}}
\newcommand{\eeqa}{\end{eqnarray*}}
\newcommand{\be}{\begin{eqnarray}}
\newcommand{\ee}{\end{eqnarray}}
\title{CSMA Local Area Networking\\ under Dynamic Altruism\thanks{The work 
was supported by NSF CISE grants 0524202 and 0915928
and by a Cisco Systems URP gift.}
}

\author{
P. Antoniadi$\mbox{s}^*$, S. Fdid$\mbox{a}^*$, 
C. Griffi$\mbox{n}^\dagger$, Y. Ji$\mbox{n}^\ddagger$, and 
G. Kesidi$\mbox{s}^\dagger$\\
*CS Dept, Univ. Pierre \& Marie Curie (LIP6), 
\{panayotis.antoniadis,serge.fdida\}@lip6.fr \\
$\dagger$ ARL, CS\&E and EE Depts, Penn State University, 
\{cxg286,gik2\}@psu.edu   \\
$\ddagger$  EE Dept, KAIST, South Korea, youngmi\_jin@kaist.ac.kr
}

\begin{document}
\maketitle

\begin{abstract}
In this paper, we consider medium access control of local area networks
(LANs) under limited-information conditions as befits a distributed system.
Rather than assuming ``by rule" conformance to a protocol designed to
regulate packet-flow rates (\eg CSMA windowing), we begin with a
non-cooperative game framework and build a dynamic altruism term into the
net utility. The effects of altruism are analyzed at Nash equilibrium for
both the ALOHA and CSMA frameworks in the quasi-stationary (fictitious
play) regime. We consider either power or throughput based costs of
networking, and the cases of identical or heterogeneous (independent)
users/players. 
In a numerical study we consider diverse players, 
and we see that the effects of altruism for similar players can be
beneficial in the presence of significant congestion, but excessive
altruism may lead to underuse of the channel when demand is low.
\end{abstract}

\section{Introduction}

Flow and congestion control are fundamental networking problems due to the
distributed, information-limited nature of the decision making process in
many popular access technologies. Various distributed 
mechanisms have been implemented to cooperatively desynchronize demand, \eg
TCP, ALOHA, CSMA.  Typically, when congestion is detected, all end-devices
are expected to slow down their transmission rates and then increase again
slowly hoping to find a fair and efficient equilibrium\footnote{RED 
\cite{Lyles99} was intended to anticipate congestion and desynchronize 
TCP backoff actions.}.
The fact that this process is not incentive compatible (a user/player 
could selfishly benefit by not
following the prescribed protocol) raises two important issues. 

First, the fact that users could have access to alternative implementations
of the prescribed (``by rule") protocols, \eg ones that slow down less than
they should, could lead to an unfair allocation or even congestion
collapse (see \eg  \cite{Cagalj05, Raya06}).  To address such threats,
there is a steadily growing literature that analyzes the equilibria of
different distributed network resource allocation {\em games}
\cite{Yaiche00, Jin02b, Basar02, Wicker03, Jin05, Cagalj05, Chiang06,
Lee07, Long07, Jin07, Cui08, Ma09}.  The theoretical models considered
ideally would allow for more informed choices in the implementation of the
corresponding flow and
congestion control protocols (\eg by associating a utility
function to end-devices, which can then be the basis of actions by
rationally selfish players). 

A ``fair" resource allocation may not be optimal from an economic point
of view. So, a second important challenge in the design of distributed
flow and
congestion control protocols is to enable users to optimize their utility
of bandwidth while simultaneously maximizing the total welfare in the
system. One simple approach is to set a pricing mechanism on resource 
consumption (or
account for the cost of networking) and thus allow users to express their
net utility through their willingness to pay.
Otherwise, there may not be a 
trustworthy way for heterogeneous users that need more to get more, thus
increasing the incentives of certain users to bypass the prescribed
protocol.

For a random-access LAN, several authors have recently considered the
problem of distributed optimization of a global objective (total
throughput, social welfare) subject to a fairness constraint.  For example,
in \cite{Cui08}, a utility function design problem is studied  considering
estimation errors of the network state.  In a  Markovian setting without
fictitious play\footnote{\ie without steady-state estimates of certain
quantities.}, \cite{Ma09} introduces a cooperation parameter (a probability
to stop transmitting), but follows a detection and punishment methodology
regarding selfish behavior.  In \cite{Heusse05}, a window-update algorithm
that tries to directly  minimize the average idle time of the channel  
is proposed.
As our main contribution, we formulate and analyze a novel CSMA medium access
control game with conditionally altruistic players.  Altruistic tactics in
evolutionary/mean-field games have long been considered, see
\cite{Malhame_CDC_2010} as a recent reference.  For example, altruism has
been modeled 
as a user's {\em statically} personalized weight on the utility of others
in games of: network formation \cite{martignon}, packet
forwarding in delay tolerant networks \cite{hui}, routing \cite{Azad09},
and medium access control by us in \cite{Kesidis10-cdc}.
See also
\cite{Hoefer09,Caragiannis10,Chen11}  for examples 
assuming a fixed set of 
altruism parameters that characterize each user or pair of users.

In the following, we consider a fictitious play model where altruism by one
user is based on perceived mean throughput of the other players
modulated (\ie made ``dynamic") by factoring the estimated mean
total channel idle time \cite{Heusse05}.
Large idle time may be a signal that competing
devices are also behaving in a socially sensitive manner, expressing
the current ``social norm," or it could simply reflect low traffic
demand.
So, we model the altruistic motivations of a user in a simpler
framework of heterogeneous users in which the user will employ
less than her ``fair" share when she doesn't really need it,
but under the constraint that others do the same.
The resulting equilibria that such
altruistic devices could reach are studied in terms of stability and under
assumptions of asynchronous/multi-rate user/players. Finally, we do not assume
that the users share information and act in a coordinated fashion, \ie so
as to form a player coalition.

Note that our system with heterogeneous users will respond similarly to a
selfish user with low throughput demand and a more altruistic one with high
throughput demand. Moreover, excessive altruism will simply result in an
underused  channel. To address this, a possible improvement would be to
include a measure on the expected congestion levels based on the number of
users sharing the same channel and to limit the altruism factor in end-user
utilities so that channel underuse does not result.  
We leave these
issues, as well as applications to other MAC scenarios not considered here,
to future work.

This paper is outlined as follows. In Section \ref{altr-sec}, we give a
brief background on altruistic behavior. A fictitious-play model with
dynamic altruism for a slotted-ALOHA LAN is given in Section
\ref{aloha-sec}.  In Section \ref{model-var-sec}, variations of the LAN
model are considered.
Numerical studies are given in Sections \ref{num-sec}
and \ref{num-sec-diversity}, the latter considers player diversity.  A
discussion of TCP flow control and congestion avoidance, in the context of
player altruism, is given in Section \ref{tcp-sec}.  Finally, in Section
\ref{summary-future-sec}, we conclude with a summary and discussion of
future work.

\section{Background on altruistic behavior}\label{altr-sec}

Economists are often criticized for the common assumption 
that humans are ``rational'' (\ie purely self-interested), 
which leads to a pessimistic view of the outcome of various 
formulated game-theoretic models.  In reality, many people act 
``altruistically'', defined as an ``unselfish concern for 
or devotion to the welfare of others'' \cite{altruism-def}.
In fact, despite this selfishness stereotype for economics, 
certain branches, such as behavioral economics and 
experimental economics, do incorporate social, cognitive,
and psychological factors in their models of human behavior
(see  \cite{loss-aversion} for a historical overview), in a way
not typically captured in  cooperative game-theoretic frameworks.
Especially since the early 1980s, there is a very rich and constantly 
growing literature on experimental economics,
whose ambitious objective is to devise (in particular, rational) 
economic models that express accurately the expected altruistic 
behavior of humans in certain settings through the analysis of 
experimental data, \eg \cite{ledyard, levitt}.

Two common scenarios in which altruistic behavior consistently
appears is in ultimatum bargaining and public-good contribution games.  
Ultimatum bargaining reveals the altruistic
instincts of humans in a resource-sharing problem in which one player
decides how to share a fixed amount of money and the other decides whether
to accept or reject sharing: here rejection leaves both with zero profit.
Experiments show that people altruistically sacrifice their own profit
to punish unfair' decisions by others. 
Analysis of more traditional public-good experiments,
where players determine their individual contribution toward the
construction of a pure public good, similarly challenges the 
assumption that free riding is always the dominant strategy. 

Of course, it is well accepted that many factors can affect human behavior, 
from the nature of the game itself to small details of the experiment 
implementation.
Nevertheless, there are numerous interesting building blocks of human
motivation that have been identified in the process and can help us 
build more realistic economic models.
For example, Andreoni \cite{andreoni_warm_glow} shows that there are
two different fundamental ways to capture altruism in a 
utility model based on whether the source of ``irrational'' 
satisfaction is the utility of the other users (satisfaction from the
benefit of others), or one's own contribution (satisfaction from being good).  

An important lesson of experimental economics 
is that altruism does not seem to be a static and hardwired 
characteristic of humans but depends on many aspects
of the environment. In other words, the level of altruism of
an individual is dynamic and could change over time depending on
the context and the behavior of the group.
Indeed, the cooperation rate 
in many experiments has been proven to be much higher if subjects 
know that there is a possibility of meeting the same partners 
again in future periods \cite{fehr_nature}, when their perception 
on the overall level of altruism in their group is high
\cite{bicchieri_experiments}, or even just by a
positive framing of the experiment \cite{andreoni_framing}.

From these and many other contextual factors that can affect the
cooperation levels in a group, social norms is perhaps the most 
influential (see \cite{ostrom_norms, bicchieri_norms}) 
but complex to incorporate in a simple economic model.
To this end, Fehr and Schmidt \cite{fehr_inequity} have proposed an
utility function explaining the altruistic behavior of people in
ultimatum experiments, which incorporates a measure of fairness (or 
``inequity aversion'') in a static way, \ie its
main parameters are indifferent to the dynamics of the 
system.
As a more realistic but less tractable alternative, H. Margolis 
argues in favor of a more dynamic and complex model, 
called ``neither selfish nor exploited'' \cite{margolis07}, which proposes 
a dual utility model which takes into account the history of one's actions, 
the current overall behavior, the effect of altruistic action, and the 
developed norms in a society.

In our scenario, the high complexity of human nature and the
surrounding social environment plays a less important role since the
cooperation game that we study is limited in time, the identity
of the players are hidden, the stakes are relatively low, and the 
decisions of users are mediated through a programmed device.
So, in our model of the following section we incorporate in a
simple utility function the effect of the external manifestation 
of altruistic behavior, that is a {\em statistical norm}
as termed in \cite{margolis07},
or simply ``what others do" \cite{bicchieri_experiments}.
To perceive this, the availability of reliable information about 
the group's statistical behavior is critical. Our use of the mean idle
time per active player
to determine the level of altruism in the system is realistic
in terms of information availability since it can be easily measured
by the different users, though, again, low demand could be 
mistakenly taken for altruistic behavior and vice 
versa, \cf Section \ref{summary-future-sec}.

When altruism can bring future concrete benefits, one could also see 
altruistic behavior as a long-term net utility optimization. 
A characteristic example is the notion of direct and indirect reciprocity 
and the related work in evolutionary game theory that tries to explain the 
source of cooperative behavior in nature \cite{Axelrod84, Nowak06}.  
We leave to future work the study of 
such evolutionary extensions of the LAN systems we formulate below,
again \cf Section \ref{summary-future-sec}.

Finally, some networking mechanisms presume cooperation ``by rule"
to affect communal benefit, such as flow and congestion control. 
Notwithstanding a tendency to altruism in the typical user, 
these mechanisms may be easily exploited by unpoliced, 
greedy individuals, \cf the discussion of Section \ref{tcp-sec}.

\section{Slotted-ALOHA random-access LAN with dynamic altruism}
\label{aloha-sec}


\subsection{Altruistic framework for with power based cost and concave
utility of throughput}

Consider a slotted ALOHA random-access LAN wherein the $N\geq 2$
participating nodes control their access probability parameter,  $q$.  A
basic assumption is that nodes' control actions are based on observations
in steady-state, \ie ``fictitious play" \cite{Brown51}, resulting in a
quasi-stationary dynamical system \cite{Jin02a,Jin02b,Wicker03} based on
the mean throughputs:
\beqa
\gamma_i(\uq) & =&  q_i \prod_{j\not=i} (1-q_j).
\eeqa
Another basic assumption in the following is that the source of a
successful transmission is evident to all other participating nodes.  We
further assume that the degree of altruism $\alpha_i$ of each node $i$
depends on the activity of the other users as:
\beqa
\alpha_i(\uq_{-i}) &  = & \prod_{j\not =i} (1-q_j) ~=~\frac{\gamma_i(\uq)}{q_i}\\
& = & \gamma_i(\uq) + \prod_j (1-q_j),
\eeqa
where the second term is just the mean idle time of the channel; thus,
every node can easily estimate  its (dynamic) altruism.  By using its
control action (strategy), $q_i$, each $i$ seeks to maximize its {\em net}
utility: 
\be
V_i(\uq)  & = &  
C_i\log(\gamma_i(\uq))  
+ A_i \alpha_i(\uq_{-i})
\overline{\gamma}_{-i}(\uq) 
- M_i q_i
\label{alt-util}
\ee
where: the dynamic altruism factor $\alpha$ modulates the 
contribution of the mean service of
all other players to the net utility of player $i$,
\be\label{gamma-bar}
\overline{\gamma}_{-i}(\uq)
		& =&   \frac{1}{N-1}\sum_{j\not=i}\gamma_j(\uq);
\ee
the utility derived by one's own throughput is modulated by a concave
function \cite{Jin02a,Jin02b,Jin05} as modeled here in the form of a 
logarithm
(for tractability); and we have assumed a power based cost\footnote{Power
based costs are borne whether or not the transmission is successful.} $M
q$. Note that because we assume that the source of each successfully
transmitted packet is evident to all nodes,  each node $i$ can easily
estimate $\overline{\gamma}_{-i}$.  Again, though each player $i$ optimizes
$V_i$ in a non-cooperative fashion, the game is called altruistic to
reflect the second term in (\ref{alt-util}).

A single-play slotted ALOHA game between two identical players is similar
to the game {\em chicken}. If $\xi<1$ is the cost of transmission
and the (normalized) payoff of successful transmission is 1, then the
following table gives net payoffs for collective action (transmit (Tx) or
not) by the players (P1,P2):\\

\begin{center}
\begin{tabular}{c|c|c}
\backslashbox{P2}{P1} & Tx & no Tx \\ \hline
Tx & $(-\xi,-\xi)$ & $(0,1-\xi)$ \\ \hline
no Tx & $(1-\xi,0)$ & $(0,0)$ 
\end{tabular}
\end{center}
~\\

The single-play game has three Nash equilibria: two ``pure" strategies,
(Tx,no-Tx) and (no-Tx,Tx), and  one mixed strategy: Tx with probability
$q^*$ (and don't Tx with probability $1-q^*$), where
$q^* = 1-\xi$ jointly minimizes the expected net
gains, $(1-\xi)q_k(1-q_{3-k}) -\xi q_k q_{3-k}$, of players $k\in\{1,2\}$.  

In the following, we consider an {\em iterated} version of this game
where players pursue mixed strategies based on observations of
throughput $\gamma_i$ observed in steady-state. 
Note that if we further assume that nodes are aware of the
$C,M$ parameters of other nodes, then we can replace $\overline{\gamma}$
with the net utility of the other players
as in \cite{Kesidis10-cdc} (particularly for
throughput based costs $M\gamma$). 

~

\begin{proposition}
\label{identical-player-game-claim}
If the  game is synchronous-play and all users $i$ have the same
(normalized) parameters 
\beqa
c:=C_i/M_i ~<~1  &\mbox{and} & a:=A_i/M_i,
\eeqa
then there is  a symmetric Nash equilibrium $\uq^* = \qstar \uone$, where
$0<\qstar<1$ is a solution to
\be\label{qstar-equ}
f(q) ~:=~ a q^2  (1-q)^{2N-3} +q-c & = & 0.
\ee
\end{proposition}

\begin{IEEEproof}
When $q_i=q$ for all $i$,  
the first-order necessary conditions of a symmetric Nash equilibrium,
\beqa
0 & = & \frac{\partial V_i}{\partial q_i} (q\uone)~ =~ -\frac{M}{q} f(q),
\eeqa
\ie equivalent to (\ref{qstar-equ}).
Note that $f(0)=-c <0$ and $f(1)=1-c>0$, the latter by hypothesis. 
So, by the continuity of $f$ and the
intermediate value theorem, a root of $f$ exists in $(0,1)$. 

All such solutions $\qstar\uone$ correspond to Nash equilibria because
$\partial^2 V_i(\uq) /\partial q_i^2 = -C_i/q_i^2 < 0$ for all $i,\uq$.
\end{IEEEproof}

~\\
The following corollary is immediate.

~\\

\begin{corollary}
There is a unique symmetric Nash equilibrium point (NEP)
if 
$\min_{q\in(0,1)}f'(q) > 0$
(\ie $f$ is strictly increasing), a condition
on parameters $N$ and $a$.  
\end{corollary}


~\\
Note that there may be non-symmetric Nash equilibria in these games,
even for the case of homogeneous users, \eg \cite{JinKes12}.
Also, it is well known that
Nash equilibria of iterative games are not necessarily asymptotically
stable, \eg \cite{Seade80,Levine85,ZhangZhang96}.  In
\cite{Jin02a,Jin02b} for a slotted-ALOHA game with throughput based costs
$M\gamma$, using a Lyapunov function for arbitrary $N\geq 2$ players, a
non-cooperative two-player ALOHA was shown  to have two different
interior\footnote{\ie not including the stable boundary deadlock
equilibrium at $\uq=\uone$.} Nash equilibria only one of which was locally asymptotically
stable (see also \cite{Menache07}). 

For  stability analysis of our altruistic
game, consider the discrete-time ($n$), synchronous-play 
gradient-ascent dynamics,
\be\label{update-rule}
q_i(n) & = &    \mbox{arg}\max_{q_i} V_i(q_i; \uq_{-i}(n-1))
~~~\forall i.
\ee
In a distributed system\footnote{\cf Section \ref{asynch-play} for a
discussion of asynchronous play.},  the corresponding continuous-time
Jacobi iteration approximation is:
\be\label{jacobi-iter-rosen}
\dot{q}_i(t) & = & 
\frac{\partial V_i}{\partial q_i}(\uq(t)) 
~~~\forall i,
\ee
and is motivated when players take small steps toward their currently optimal
response, \ie better-response dynamics \cite{Shamma05}.
That is, for positive step-size $\varepsilon\ll 1$
(\ref{jacobi-iter-rosen}) approximates the discrete-time better-response
dynamics,
\be
q_i(n) & = & q_i(n-1)+\varepsilon 
\frac{\partial V_i}{\partial q_i}(\uq(n)) 
~~~\forall i, \label{beji}
\ee
which is a kind of distributed gradient ascent.  The Jacobi iteration is
also motivated by the desire to take small steps to avoid regions of
attraction of undesirable boundary NEPs, particularly  those corresponding
to the capture strategy ($q_i=1$ for  some $i$). Note that when more than
one player selects this strategy, the result is a bad outcome for the game
{\em chicken} or a deadlocked ``tragedy of the commons." Additionally the
players avoid the opt-out strategy ($q_i=0$ for some $i$).  In summary,
(\ref{beji}) represents a repeated  game in which players adjust their
transmission parameters $q_i$ to (locally) maximize their net utility
$V_i$. 

To find conditions on the parameters of net utilities (\ref{alt-util}) for
local stability of the equilibria, we can apply the Hartman-Grobman theorem
\cite{Perko01}
to (\ref{jacobi-iter-rosen}), \ie check that the Jacobian is negative
definite.  The following proposition uses the conditions of \cite{Rosen65}
for stability (and uniqueness) for a special case.

~\\

\begin{proposition}\label{stability-claim}
In the case where players have the same parameters $C$ and $A$, the
symmetric NEP $\qstar \uone$ is locally asymptotically stable under the
dynamics in (\ref{jacobi-iter-rosen}) when the normalized parameters satisfy
\be\label{stability-cond}
C & > & 2(N-1)A.
\ee
\end{proposition}

\begin{IEEEproof}
By \cite{Rosen65}, the result follows if
the symmetric $N\times N$ matrix $H(\uq)$ is negative
definite, where 
\beqa
H_{ij} & = &  
\frac{\partial^2 V_i}{\partial q_i\partial q_j}
+\frac{\partial^2 V_j}{\partial q_j\partial q_i}.
\eeqa
First note that, for all $i$, 
\beqa
H_{ii}(\uq) & = & -\frac{C}{q_i^2} ~<~ -C.
\eeqa
For $l\not= i$,
\beqa
\frac{\partial^2 V_i}{\partial q_i\partial q_l} 
& = & \frac{\partial}{\partial q_l} \left(\frac{C}{q_i}
-A\alpha(\uq_{-i})
\tfrac{1}{N-1}\sum_{j\not=i}
q_j\prod_{k\not =i,j } (1-q_k) 
\right)\\
& = &  
A\prod_{j\not =i,l } (1-q_j)
\tfrac{1}{N-1}\sum_{j\not=i}
q_j\prod_{k\not =i,j } (1-q_k) \\
&  & ~
+A\alpha(\uq_{-i})\tfrac{1}{N-1}\left(
\sum_{j\not = i,l} q_j
\prod_{k\not =i,j,l } (1-q_k)  \right.\\
& & ~~~~~~\left.
-\prod_{k\not =i,l } (1-q_k) 
\right).
\eeqa
Now because
$0<q_i<1$ for all $i$ and the triangle inequality,
\beqa
|H_{ij}(\uq)| & \leq  & 2A ~~ \forall j\not=i.
\eeqa
So, by the
Gershgorin circle (disc) theorem (see p. 344 of \cite{horn0}),
all of $H(\uq)$'s eigenvalues are less than $-C + (N-1)2A$.
So, if (\ref{stability-cond}) holds, then
all the eigenvalues of $H(\uq)$ are negative,
and so $H(\uq)$  is  negative definite.
\end{IEEEproof}


%
%
%
%
%
%

\subsection{The marginal effect of altruism}


In this section, we will write $\qstar$ (of the symmetric NEP $\qstar\uone$
in symmetric  users case) as a function of the normalized altruism
parameter $a:=A/M$, $\qstar(a)$.  Note that $\qstar(0)=c:=C/M$.  

Recall that the total throughput for slotted ALOHA, $Nc(1-c)^{N-1}$, is
maximal when $c=1/N$. The maximum total throughput is $(1-1/N)^{N-1}
\approx \e^{-1}$ for large $N$, \ie the maximum throughput per player is
$1/(N\e)$ in this {\em cooperative} setting {\em without} networking
costs.

So, if $c> 1/N$, \ie total throughput is less than $\e^{-1}$ because of
excessive demand (overloaded system), then a marginal increase in altruism
from zero ($0<a\ll 1$) will  cause a marginal decrease in $\qstar\downarrow
1/N$, resulting in an increase in throughput per user $\gamma \uparrow
1/(N\e)$.  Also, if $c< 1/N$, \ie total throughput is less than $\e^{-1}$
because of a lack of demand (an underloaded system), then a marginal increase
in altruism from zero  will  again cause a marginal decrease in $\qstar$,
but here  resulting in a decrease in throughput $\gamma$ (further away from
the optimum $\e^{-1}$).  See Section \ref{anarchy-sec} below.

\section{Model variations}\label{model-var-sec}

In this section, we discuss model variations, which we subsequently
analyze. An analytically
straightforward variation is to simply  use $(N-1)\overline{\gamma}$
(\ie just mean {\em total} channel idle-time),
instead of $\overline{\gamma}$ 
given by (\ref{gamma-bar}),
thus not requiring an estimate of the
the number of active users, $N$.  However, 
the mean idle-time {\em per active user} $\overline{\gamma}$ 
better captures the current levels of altruistic behavior
in the system and is more
consistent with ideas of inequity aversion \cite{fehr_inequity}.
So we will not explore this simple variation further here as, again,
we are herein neither interested in maximizing social welfare nor related 
``efficiency" issues
captured by a global criterion (accommodating
users with very different magnitudes of demand in a typically
additive way); rather, we are interested
in the effect of altruism on distributed, non-cooperative network-access games.
More ambitious  model variations than those discussed in this 
section  are mentioned in
the concluding Section \ref{summary-future-sec}.

Note how our model of altruism leads to neither selflessness nor full
cooperation,  but is closer to a (rationally) selfish model, \ie  
again
as H. Margolis characterized it, 
neither selfish nor exploited \cite{margolis07}.
Also in our model,
altruism needs to accommodate the limited and potentially unreliable
information of a distributed network.




\subsection{Throughput based costs}\label{throughput-costs}

In \cite{Kesidis10-cdc} we considered throughput based costs
with a static altruism parameter and with utility proportional to
throughput.  Instead of (\ref{alt-util}), for throughput based costs with
dynamic altruism and utility being a concave (log) function of throughput,
we can model the net utility as:
\be
\tilde{V}_i(\uq)   =   
C_i\log(\gamma_i(\uq))  + A_i \alpha_i(\uq_{-i})
\overline{\gamma}_{-i}(\uq) - M_i \gamma_i(\uq).
\label{alt-util2}
\ee

Proposition \ref{identical-player-game-claim} can easily be
adapted for power based
costs.  Instead of (\ref{qstar-equ}), the
first-order necessary condition for a symmetric Nash equilibrium $q\uone$
under throughput based cost  is
\be\label{qstar-equ2}
\tilde{f}(q) ~:= a q^2  (1-q)^{2N-3} +q(1-q)^{N-1}-c & = & 0.
\ee
All solutions $q$ for (\ref{qstar-equ2}) correspond to NEPs $q\uone$ because
$\partial^2 \tilde{V}_i(\uq) /\partial q_i^2 = -C_i/q_i^2 < 0$ for all $i,\uq$ 
(as for power based cost).
Note that 
$\tilde{f}(0)=\tilde{f}(1)=-c<0$, so we cannot simply use the intermediate
value theorem as we did for Proposition \ref{identical-player-game-claim}
to establish existence of a symmetric Nash equilibrium when $c<1$.  Here,
existence requires 
\be\label{throughput-NEP-condition}
\max_{0<q<1} \tilde{f}(q) & \geq & 0,
\ee
a condition on $N,c,a$.
Note that if the inequality in (\ref{throughput-NEP-condition}) strictly
holds then there will be an even number of symmetric NEPs, again by the
intermediate value theorem. If the maximum equals zero then there may be a
unique symmetric NEP.

\subsection{Proportional throughput utility}

Suppose that utility is simply proportional to throughput and cost is power
based, \ie the net utility is
\be
\hat{V}_i(\uq)   =   
C_i\gamma_i(\uq)  + A_i \alpha_i(\uq_{-i})
\overline{\gamma}_{-i}(\uq) - M_i q_i.
\label{alt-util3}
\ee
Note that  the  net utility $\hat{V}_i$ is {\em linear} in $q_i$ (this
would also be the case if throughput based costs were involved).  This
normally leads to candidate ``bang-bang" Nash equilibrium play-actions,
$q_i\in\{0,1\}$ for all players $i$; \ie the players are either out of the
game ($q_i=0$ if $\partial \hat{V}_i/\partial q_i <0$) or are {\em all in}
($q_i=1$ if $\partial \hat{V}_i/\partial q_i>0$).  Note that  the latter
play action, potentially leading to the deadlock of ``tragedy of the
commons", is {\em not} an equilibrium here because if $q_j =1$ then
$\partial \hat{V}_i/\partial q_i = -M < 0$ for all $i\not= j$.

It turns out that for  this case, there is a symmetric interior equilibrium
$q\uone$ for the identical players case with $0<q<1$, \ie where
\be
\hat{f}(q) & := & \tfrac{\partial \hat{V}_i}{\partial q_i}(q\uone) \nonumber\\
& = &  c(1-q)^{N-1}-aq(1-q)^{2N-3}-1   ~ = ~  0.
\label{qstar-equ3}
\ee
If $c>1$, $\hat{f}(0)=c-1>0$ and $\hat{f}(1) =-1<0$ and so there is a
solution to $\hat{f}(q)=0$ for $0<q<1$ by the intermediate value theorem.
It should be noted, however, that such an interior Nash equilibrium
$q\uone$ is not stable, \ie it's a saddle  point in the domain $[0,1]^N$.

\subsection{Asynchronous/Multirate Players}\label{asynch-play}

Asynchronous players were considered previously in \cite{Jin05}
using the ideas from \cite{Basar95, Bertsekas89}. A very similar approach
can be used to extend the results herein to account for the effects of
asynchronous play.  Numerical results for this case are given in Section
\ref{asynch-play-numerical} below.


\section{Numerical studies for identical players at Nash equilibrium}\label{num-sec}

\subsection{Power based costs}\label{numerical-throughput}
\begin{figure}[ht]
\centering
\subfigure[Ranging $N$]{
\includegraphics[width=3.5in]{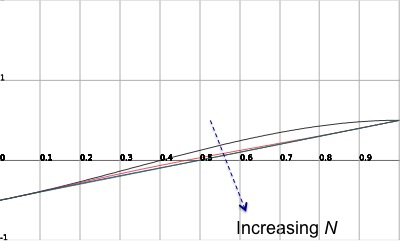}
\label{fig:Power_Nrange}
}
\subfigure[Ranging $a$]{
\includegraphics[width=3.5in]{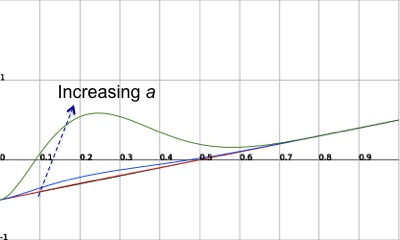}
\label{fig:Power_arange}
}
\caption{Power based costs}\label{fig:Power}
\end{figure}

For normalized utility  parameter $c=0.5$ and normalized altruism parameter
$a=1$, Figure \ref{fig:Power_Nrange} is a plot
of $f$ in (\ref{qstar-equ}); \ie for power
based costs, for $N=2,3,5,10$ players. The root at $q=0.4$ corresponds to
$N=2$ (\ie corresponding to NEP $q\uone$)  and, as the first term of $f$
becomes negligible, the root at $\approx 0.5$ corresponds to the $N>2$ cases.
For $c=0.5$ and $N=5$, Figure \ref{fig:Power_arange} is a plot of $f$ for
$a=0.1,1,10,100$.  Note that $a=100$ corresponds to the larger curve which has
a the smaller root $q$, \ie under ``excessive" altruism the NEP
$q\rightarrow 0$.

\subsection{Throughput based costs}
\begin{figure}[ht]
\centering
\subfigure[Ranging $N$]{
\includegraphics[width=3.5in]{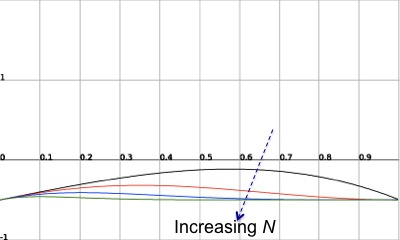}
\label{fig:Throughput_Nrange}
}
\subfigure[Ranging $a$]{
\includegraphics[width=3.5in]{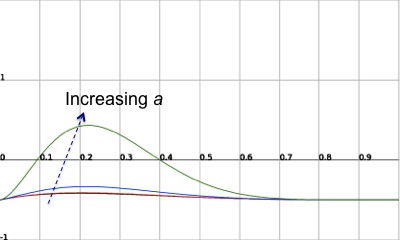}
\label{fig:Throughput_arange}
}
\caption{Throughput based costs}\label{fig:Throughput}
\end{figure}

Using the same parameter cases as those for power based costs, Figure
\ref{fig:Throughput_Nrange} is a plot of $\tilde{f}$ in (\ref{qstar-equ2})
for $c=0.5$, $a=1$ and $N=2,3,5,10$.  Figure \ref{fig:Throughput_arange} is a
plot of $\tilde{f}$ in (\ref{qstar-equ2}) again for $c=0.5$, $N=5$ and
$a=0.1,1,10,100$.  The larger curve, corresponding to $a=100$, has two
zero-crossings $q$ at approximately 0.1 and 0.4, \ie has two different
symmetric NEPs $q\uone$.  The other parameter sets do not possess an
interior NEP, a situation which will be remedied if we reduce the utility
$c$ from 0.5 to zero; that is, increasing $\tilde{f}$.

\subsection{Throughput-proportional utilities and costs}
\begin{figure}[ht]
\centering
\subfigure[Ranging $N$]{
\includegraphics[width=3.5in]{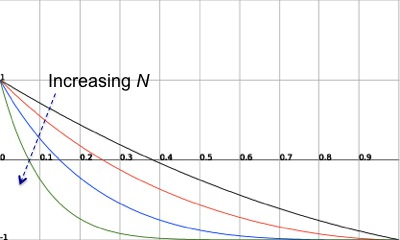}
\label{fig:Proputil_Nrange}
}
\subfigure[Ranging $a$]{
\includegraphics[width=3.5in]{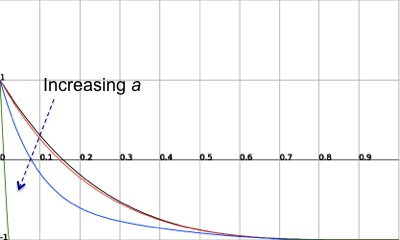}
\label{fig:Proputil_arange}
}
\caption{Throughput proportional utility and cost}\label{fig:Proputil}
\end{figure}

Figure \ref{fig:Proputil_Nrange} is a plot of $\hat{f}$ in
(\ref{qstar-equ3}) for $c=2$, $a=1$ and $N=2,3,5,10$.  Figure
\ref{fig:Proputil_arange} is a plot of $\hat{f}$ for $c=2$, $N=5$ and
$a=0.1,1,10,100$.  Following intuition, the lower curves (and lower roots,
NEPs) correspond to larger $N$ (larger congestion leading to lower
throughput) or larger $a$ (greater altruism again leading to lower
throughput), in a monotonic fashion when all other parameters fixed.

\subsection{An example comparing altruism and non-cooperation}\label{anarchy-sec}

In this section, we compare the Nash equilibria under altruistic player
action with equilibria in purely non-cooperative scenarios.  For all
scenarios, we considered the case of power based costs, log-utility of
throughput, normalized utility parameter $c =0.5$, and identical users.
For the purely non-cooperative scenario, \ie $a=0$, the symmetric Nash
equilibrium $q=c=0.5$ is simply obtained by solving (\ref{qstar-equ}).  For
the scenarios with altruism,  the normalized altruism parameter was taken
to be $a=20$.  Recall that for static altruism, $\alpha\equiv 1$.  At Nash
equilibrium $\uq^*=q^*\uone$, the throughput ($\gamma^*=q^*(1-q^*)^{N-1}$)
and utility (\ref{alt-util}) performance per user is given in the following
table, in decreasing order of throughput.




\begin{tabular}{c|c|c|c|c}
Scenario & $N$ & $q^*$ & $\gamma^*$  &  $V^*/M$ \\ \hline
Dynamic Altruism & 4 & .22 & .1044 & -0.36 \\
Static Altruism & 4 & .16 & .0935 &  0.53 \\
Non-cooperative   & 4 & .50 & .0625 & -1.89 \\
Static Altruism & 8 & .28 & .0277 & -1.52 \\
Dynamic Altruism & 8 & .50 & .0039 & -3.27 \\
Non-cooperative   & 8 & .50 & .0039 & -3.27 
\end{tabular}


~\\
Note that for $N=8$, the altruistic component of utility at Nash
equilibrium is negligible, when comparing dynamic altruism versus
non-cooperation, owing to high contention under the assumed parameters when
there are eight users. We see that the non-cooperative scenario has poorest
throughput performance in the above examples. However, if the level of
altruism is too high, under either the static or dynamics mechanisms, the
channel may be underused; in this case, the altruism parameters could
be adjusted via an ``evolutionary" process to avoid channel
underuse.

\section{Numerical studies with player diversity}\label{num-sec-diversity}

\subsection{Players with different altruism parameters}


Consider the game with power based costs.
In this section, we consider players with different normalized altruism
parameters $a$ 
for $N=3$ otherwise identical players with normalized
parameter $c=0.5$ associated with power-based cost.  Specifically, the
first player has  $a_1$ ranging from 30 to 70, while the other two players
both have $a=50$. Note that changing $a$ in this manner will result in changes
in the NEP $q^*$ and  corresponding throughputs $\gamma^*$ (and utilities
$V^*$ per user, as shown in the following table):
\beqa
\begin{array}{c|c|c|c}
a_1 & q_1^*,q_2^*=q_3^*  & \gamma^*_1,\gamma^*_2=\gamma^*_3 & 
V^*_1, V_2^*=V_3^*  \\ \hline
30 &  0.15,0.10 &  0.13,0.074   & 0.754,2.37 \\ 
40 & 0.12,0.10 & 0.10,0.080  &  1.40, 2.24\\ 
50 & 0.10,0.10 & 0.083,0.083   & 2.10,2.10  \\ 
60 &  0.091,0.11 & 0.073,0.087  &  2.79, 1.83\\ 
70 & 0.079,0.11 & 0.063,0.090   & 3.56, 1.82
\end{array}
\eeqa
Clearly, increased altruism, $a_1>50$, by player 1 resulted in lower
throughput  for him and higher throughput for the other two players.
Similarly, decreased altruism by player 1, $a_1<50$, resulted in higher
throughput for him and lower throughput for the other players.

%


\subsection{Sizes of  regions of attractions  under different 
play-rates}\label{asynch-play-numerical}



\begin{figure}
\centering
\includegraphics[width=3.5in]{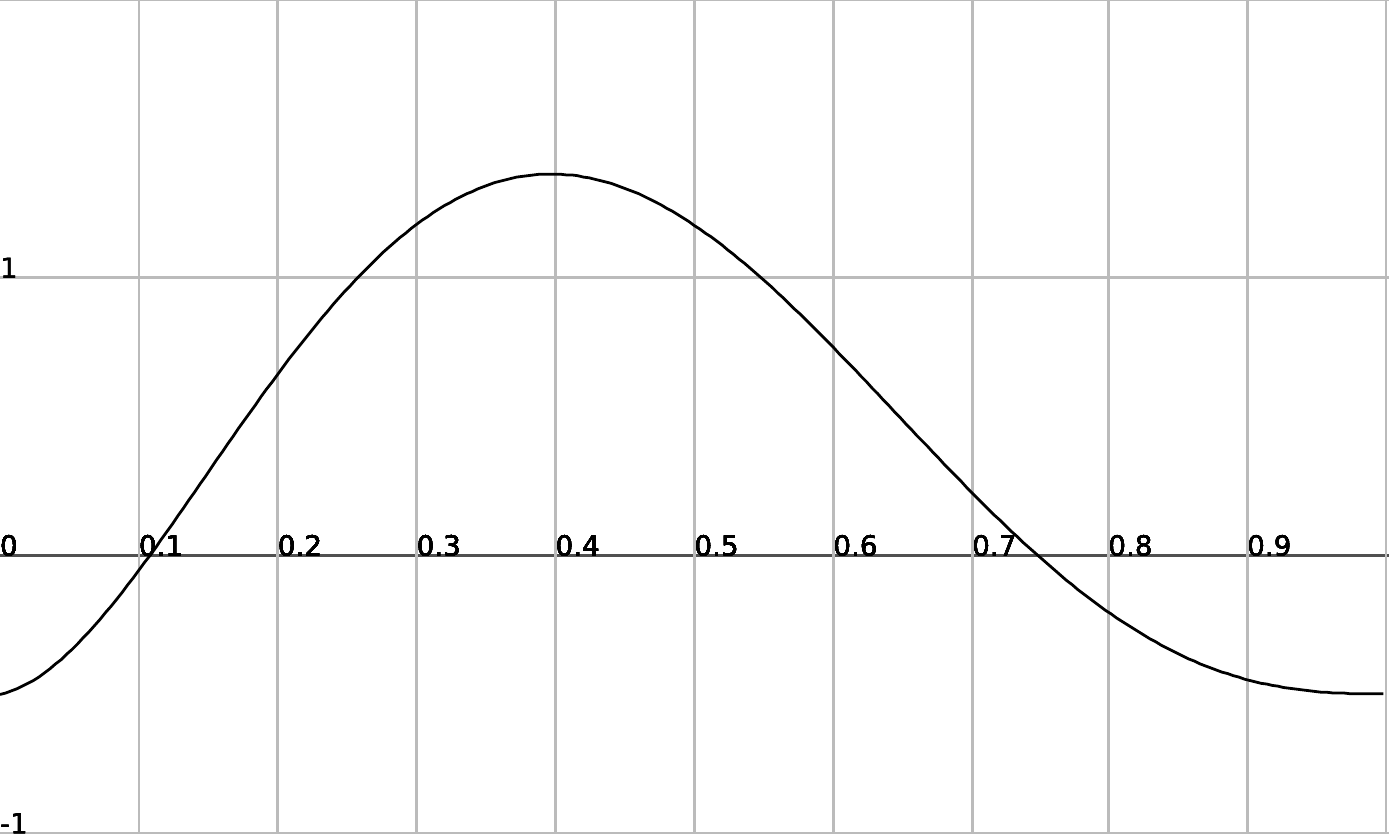}
\caption{Power based costs with  $N=3$, $a=50$ and $c=0.5$}\label{fig:attraction}
\end{figure}

In this section, we study how the volume of the regions of attractions of
different equilibria are sensitive to players adopting different
play-rates, while retaining our assumption of fictitious/quasistationary
play.  Consider the case of $N=3$ players two of whom have the same play rate
while the other adopts a play rate  that is a multiple, $r$, of the other
two. We consider the case of throughput based costs as in Figure
\ref{fig:Throughput_arange}.  That is, 
\be\label{attraction-dynamics}
\dot{\uq}_i(t) & = & 
r_i\frac{\partial \tilde{V}_i}{\partial q_i}(\uq(t)) 
~~~\forall i,
\ee
where $r_i=r$ for player $i=1$, otherwise $r_i=1$ and $\tilde{V}$ is given
in Section \ref{throughput-costs}.  Numerically simulating
(\ref{attraction-dynamics}) from different initial points  chosen from a
grid in the hypercube $[0,1]^3$, we counted the number of initial points
converging to a given NEP so as to estimate the volume of its region of
attraction.  Note that the introduction of such play-rate parameters $r_i$
does not change the position of the NEPs.  Using normalized parameters
$a=50$ and $c=0.5$, the function $\tilde{f}$ whose roots are the NEPs is
depicted in Figure \ref{fig:attraction}.  As can be seen from the following
table, the region of attraction is very sensitive to $r$ in the range $0.1$
to $10$.
\beqa
\begin{array}{c|c|c|}
\mbox{Volume} & \mbox{NEP}= (0.1)\uone &  \mbox{NEP}= (0.75)\uone \\
\hline
r=0.1 & 0.502  & 0.498 \\ 
r=0.25 & 0.507  & 0.493 \\
r=1 & 0.556 & 0.444 \\
r=4 & 0.839 & 0.161 \\ 
r=10 & 0.841 & 0.159  
\end{array}
\eeqa
The results are intuitive: a lower $r$ effectively corresponds to a
reluctance to be altruistic and thereby results in a smaller domain of
attraction for the more altruistic Pareto equilibrium $(0.1)\uone$
(corresponding to throughputs of $\gamma=0.081$ and utilities $V=1.94$,
respectively compared to $\gamma=0.047$ and $V=-1.43$ corresponding to
$(0.75)\uone$).

\section{Discussion: end-to-end TCP flow and congestion control}\label{tcp-sec}

Interesting game theoretic models for end-to-end window
flow control (as in TCP), based on internal bandwidth
bottlenecks\footnote{Typically with associated memory to buffer packets in
times of congestion.} have been extensively studied, see, \eg
\cite{Mo00,Shenker02}, including modifying backoff parameters.
Player diversity depends in part on differences in round-trip times (RTTs)
which correspond to responsiveness -- players with smallest RTT to a
congested bottleneck who follow traditional distributed TCP congestion
control will be the most altruistic  in that they will back-off first.
The equilibrium is a ``water-filling" where the player with largest RTT and
demand $D_1$ receives $d_1=\min\{D_1,C\}$, where $C$ is the bottleneck
capacity. The player with second largest RTT and demand $D_2$ receives
$d_2= \min\{D_2,\max\{C-D_1,0\}\}$, and similarly $d_3=
\min\{D_3,\max\{C-D_1-D_2,0\}\}$, \etc

Players can reduce the level of their  altruism by artificially delaying
their response to congestion (\ie beyond the minimal response time governed
by their RTT), in the limit being completely nonresponsive to congestion.
Though such manipulation may be more straightforward for end-users or
individual applications compared to modifying congestion backoff in layer-2
MAC protocols, a difficulty here is that there may be little knowledge
about the behavior of other players (in this highly distributed setting) to
form the basis of altruistic action; indeed, it is likely that many
players/sessions are only minimally responsive to congestion, \eg streaming
media over UDP running RTCP and p2p file-sharing using utorrent clients.

\section{Summary and  future work}\label{summary-future-sec}

In this paper, we extended a non-cooperative game framework for
information-limited MAC of a LAN by adding an altruism term that depended
on both the mean throughput of the other players and the mean channel 
idle time. The cases of heterogeneous or homogeneous users, and
of power or throughput based costs, were considered for a quasi-stationary
model of the game. A numerical study compares the per-user throughput
under dynamic and static altruism with that of purely non-cooperative
dynamics, and demonstrates the advantage of altruism under moderate levels
of congestion (number of players) in the homogeneous-player setting, and
for a heterogeneous user scenario.

In the future, we plan to depart from ideal quasi-stationary dynamics and
consider the effects of measurement error (as in \cite{Cui08,Menache07}),
leading to a more ``stochastic" version of the games considered here.  We
also plan to extend our study of dynamic altruism to the simple
power-control based medium access considered in \cite{Kesidis10-cdc} for
static altruism.  

Finally, we will consider an evolutionary ``wrapper" about the dynamics
considered here or other factors affecting the parameters  of the net
utilities (\ref{alt-util}), \eg the desire to avoid underutilization of the
channel or to account for multiple-priority transmission.  Indeed,
motivated by the prospects of significant reward, defections  by  a few
individual players may cause an evolutionary/slow-cycle of transitions
between full cooperation (limited by information availability and
associated costs), to complete non-cooperation,
via ``intermediate" altruistic behavior \cite{Nowak06}.
So, the altruism considered herein may, in principle, be  rationally
motivated based on long-term reward considerations.

\bibliographystyle{plain}

\end{document}